\documentstyle[12pt,epsbox]{article}
\baselineskip=\normalbaselineskip

\begin{document}

\begin{titlepage}

\begin{flushright}
KUCP-0091 \\ 
February 1996 
\end{flushright}
\vskip 1cm

\begin{center}
{\Large  
 Comment on Entropy in Two-dimensional Dilaton Gravity
}\vskip 2cm

{\large  
 Kouichirou Hirata
}\vskip 0.7cm

{\sl  
Graduate School of Human and Environmental Studies,\\
Kyoto University, Kyoto 606, Japan}
\vskip 1.5cm
{\large  
Yoshi Fujiwara
}\vskip 0.7cm

{\sl  
Yukawa Institute for Theoretical Physics,\\
Kyoto University, Kyoto 606, Japan}
\vskip 1.5cm

{\large  
Jiro Soda
}\vskip 0.7cm

{\sl  
Department of Fundamental Sciences,
FIHS,\\
Kyoto University, Kyoto 606, Japan}
\end{center}
\vskip 2cm

\begin{abstract}
The thermodynamic second law in the evaporating black hole space-time
is examined in the context of two-dimensional dilaton black hole.
The dynamical evolution of entropy is investigated by using
the analytical perturbation method and the numerical method.
It is shown that the thermodynamic second law holds in the 
vicinity of infalling shell 
and in the region far from the infalling shell 
in the case of semi-classical CGHS model.
The analysis of the general models is also presented and its
implication is discussed.
  
\end{abstract}
\end{titlepage}


\setlength{\normalbaselineskip}{20pt plus 0.2pt minus 01.pt}
\baselineskip=14pt

\section{Introduction}
\label{sec:introduction}

At present, the thorough understanding of quantum gravity has not been
achieved yet. To reveal the nature of quantum gravity,
it is important to figure out the quantum aspects of black hole.
The existence of the non-trivial causal space-time structure
such as black holes gives a profound insight into the character
of gravity.
Curiously, the geometry of black hole is intimately related to
thermodynamics\cite{Hawking73}.
The discovery of Hawking radiation stressed
the role of the black hole thermodynamics\cite{Hawking75}.
It also gives the picture of an evaporating black holes.
Recently, the problem of information loss for the evaporating
black hole has been discussed widely.
Originally, the thermodynamic laws of black hole were proved
in a stationary situation.
Hawking's calculation was performed on a fixed stationary black
hole.
However, Hawking radiation naturally leads us to the evaporating
black hole picture. Hence, it is legitimate to ask how
the thermodynamic laws are modified. 

In this paper, we shall investigate the thermodynamic
second law in the context of two-dimensional dilaton gravity.
The two-dimensional CGHS model \cite{CGHS} 
provides an interesting toy-model
for the study of black hole thermodynamics. For this model,
it is easy to give the classical general solution.
There exists a black hole solution.
Moreover, the thermodynamic laws hold in this model\cite{frolov}
and the standard calculation yields the Hawking radiation
with a certain temperature.
It was shown that the effective action in which the backreaction
of the Hawking radiations is included gives the evaporating
black hole picture.
On the other hand, 
Wald presented a new way to look at the entropy of black hole as a
Noether charge associated with diffeomorphism invariance of a
theory\cite{Wald93}\cite{JKM}.
In particular, the first law of black hole mechanics is a
general relation between \lq\lq boundary terms\rq\rq at event horizon
and spatial infinity in a stationary asymptotically flat
space-time.
Iyer and Wald derives an inductive algorithm for
general diffeomorphism invariant theories\cite{IyerWald},
and the expression of the entropy is given by
\begin{eqnarray}
  \label{entwald}
  S = \frac{4\pi}{\sqrt{-g}}\frac{\partial L}{\partial R},
\end{eqnarray}
in the context of two-dimensional dilaton gravity.
Here $L(g_{ab}, R)$ is the Lagrangian of the theory
and $g_{ab}$ is the metric of the spacetime and $R$ is the
scalar curvature.
Myers has proposed to use the above entropy expression in dynamical
situations and has shown that the entropy satisfies the thermodynamic
second law in the classical and semi-classical RST model
\cite{Myers94}\cite{RST92b}.
However, RST model is too special to conclude that
the thermodynamic second law holds in general.
The question we would like to address is whether the Myers's result
can be extended to unsoluble semi-classical CGHS model and the more
general models.


\section{Entropy of semi-classical CGHS model}
\label{sec:entropy_of_semi-classical_cghs_model}

\subsection{Classical model and its solution}
\label{sec:classical_model_and_its_solution}

To make this paper self-contained,
we start from classical CGHS model.
The classical CGHS action is
\begin{eqnarray}
  \label{classical_cghs_action}
  I = \frac{1}{2\pi}\int d^2x \sqrt{-g}\{e^{-2\phi}[R+4(\nabla\phi)^2
  +4\lambda^2] -\frac{1}{2}\sum^{N}_{i=1}(\nabla f_{i})^2\}
\end{eqnarray}
where $\phi$ is dilaton field, $R$ is 2d Ricci scalar,
$\lambda$ is a positive constant, $\nabla$ is the covariant
derivative associated with 2d metric $g_{\mu\nu}$,
and $f_{i}$ are N massless scalar fields.
In conformal gauge, 
\[ g_{+-} = -\frac{1}{2}e^{2\rho},\;\; g_{++} = g_{--} = 0,\;\;
 x^{\pm} = t \pm r,\]
equations of motion for $\rho$, $\phi$, $f_{i}$ are
\begin{eqnarray}  
e^{-2\phi}(2\,\partial_{+}\partial_{-}\phi
           -4\partial_{+}\phi\partial_{-}\,\phi
           -\lambda^{2}e^{2\rho}) &=& 0, \\
 -4\,\partial_{+}\partial_{-}\,\phi +4\,\partial_{+}\phi\partial_{-}\phi
   +2\,\partial_{+}\partial_{-}\rho + \lambda^{2}e^{2\rho}&=& 0, \\ 
 \partial_{+}\partial_{-} f_{i} &=& 0\;\;\mbox{(for each $i$)},
\end{eqnarray}
with the constraints
\begin{equation}
  e^{-2\phi}(4\partial_{\pm}\rho\,\partial_{\pm}\phi
   -2\partial{}_{\pm}^{2}\phi) + \frac{1}{2}\sum^{N}_{i=1}
   \partial_{\pm}f_{i}\,\partial_{\pm}f_{i} = 0.
\end{equation}
As we can take $\phi=\rho$ gauge for this model,
exact solution of this model can be obtained easily.
Vacuum solution of this model in $\phi = \rho$ gauge is
\begin{equation}
e^{-2\rho} = e^{-2\phi} = M - \lambda^{2}x^{+}x^{-}.
\end{equation}
It can be easily shown that the Penrose diagram of this space time
is the same as that of Schwarzschild's spacetime.
This fact shows that CGHS model has a black hole solution.
When $M=0$,
the solution becomes
\begin{eqnarray}
  \label{vacuum_solution}
  e^{-2\phi} = e^{-2\rho} = -\lambda^{2}x^{+}x^{-},
\end{eqnarray}
and we have the flat space time
\begin{eqnarray}
  \label{LDV}
  ds^{2} &=& -\frac{dx^{+}dx^{-}}{\lambda^{2}x^{+}x^{-}}\\
         &=& -d\sigma^{+}d\sigma^{-},
\end{eqnarray}
where we have introduced a coordinate system by
$\lambda x^{\pm} = \pm e^{\lambda \sigma^{\pm}}.$
The solution is called linear dilaton vacuum(LDV).

Let us consider the matter falling along the null line,
$x^{+} = x^{+}_{0}$,
then the energy-momentum tensor of the scalar fields becomes
\begin{eqnarray*}
  \frac{1}{2}\partial_{+}f_{i}\partial_{+}f_{i} 
  = \frac{M}{\lambda x^{+}_{0}}\delta(x^{+}-x^{+}_{0}),\\
  \frac{1}{2}\partial_{-}f_{i}\partial_{-}f_{i} = 
  \frac{1}{2}\partial_{+}f_{i}\partial_{-}f_{i}  = 0.
\end{eqnarray*}
The solution for this is
\begin{eqnarray}
  \label{classical_solution_with_infalling_matter}
 e^{-2\phi} = e^{-2\rho}
 = -\frac{M}{\lambda x^{+}_{0}}(x^{+} - x^{+}_{0})
 \theta(x^{+} - x^{+}_{0}) -\lambda^{2}x^{+}x^{-},
\end{eqnarray}
where $M$ can be shown to be the ADM mass of black hole.

\subsection{Evolution of entropy}
\label{sec:evolution_of_entropy}

\subsubsection{Entropy formulation}
Now we turn to the quantum theory.
Though we cannot perform the full quantization of gravity,
we can calculate matter contribution and obtain effective action.
In covariant non-local form, quantum correction which is called
Liouville action is
\begin{eqnarray}
  \label{liouville_action}
  I_{1} = -\frac{\kappa}{8\pi}\int d^{2}x \sqrt{-g}R
  \frac{1}{\nabla^{2}}R
\end{eqnarray}
where $\kappa = N/12$, $N$ is number of scalar fields.

Using Wald's scheme mentioned in the Introduction,
the contribution to entropy from the classical action is
\begin{eqnarray}
  \label{entropy0}
  S_{0} = 2e^{-2\phi},
\end{eqnarray}
where the right hand side is evaluated on the horizon.
As for the non-local action, careful argument is necessary
in the Noether charge analysis.
Myers showed that such a non-local term actually can be
handled \cite{Myers94}.
The result is 
  \begin{equation}
  S_1= -\kappa\int d^2 y\sqrt{-g(y)}\,G(x_H,y) R(y)
  \label{entropy1}
  \end{equation}
where $G(x,y)$ is the Green function, that is,
  \begin{equation}
  \nabla^2_x G(x,y)=\delta^2(x,y)
  \label{green}
  \end{equation}
We have to specify the boundary condition for the Green function
$G(x,y)$ in order to make the expression (\ref{entropy1}) have
a definite value. 
Combining (\ref{entropy0}) and (\ref{entropy1}), we obtain
\begin{eqnarray}
  \label{entropy_CGHS}
  S = 2e^{-2\phi} +2\kappa\rho +\kappa\log(-\lambda^{2}x^{+}x^{-}),
\end{eqnarray}
in Kruskal coordinate, and the last extra term came from boundary
condition of (\ref{green}).

The entropy (\ref{entropy_CGHS}) has a local geometrical expression
and is to be evaluated at a certain part in spacetime.
In stationary spacetime, 
it is the killing horizon.
In dynamical situation, on the other hand,
it seems to be reasonable to consider the entropy
on apparent horizon since it is physically meaningful
and a locally-determined structure formed in the
dynamical spacetime.
Following this proposal, we address the question of the
validity of the entropy increase along the
apparent horizon in this paper.

Surely we have to know the solutions of $\phi$ and $\rho$ in order to
evaluate above entropy formula.
Unfortunately, however,
we cannot solve the equations of motion in the semi-classical theory.
Thus we analyze the behavior of entropy in the 
vicinity of the infalling shell
at which apparent horizon starts.
After that, some results of numerical calculations will be
presented because the above analytic calculation cannot give the
behavior of entropy far from the infalling shell.

\subsubsection{Analytic approach}
\label{sec:analytic_approach_cghs}
We would like to evaluate the entropy along the
apparent horizon.
\begin{figure}[t]
  \begin{center}
    \leavevmode
    \psbox[width=10cm]{fig1}
    \begin{small}
    \caption{This figure shows schematic explanation of situation considered here.}
    \end{small}
    \label{fig:situation}
  \end{center}
\end{figure}
Fortunately, if we restrict our consideration to the
vicinity of the infalling shell (Figure 1.),
one can calculate the variation rate of the entropy
along the apparent horizon as
\begin{eqnarray}
  \label{evolution_of_entropy}
  \frac{dS}{d x^{+}} &=& \partial_{+}S 
  + \frac{dx^{-}}{dx^{+}}\Big|_{\mbox{\tiny A.H.}}\partial_{-}S \\
  &=& 2\kappa\,\partial_{+}\rho 
     + \frac{\kappa}{x^{+}_{0}}
     -2\lambda^{2}x^{+}_{0}
     \frac{dx^{-}}{dx^{+}}\Big|_{\mbox{\tiny A.H.}},
\end{eqnarray}
where we used the condition for the
apparent horizon ($\partial_{+}\phi=0$),
and all of the quantities are evaluated at the shell\cite{tada}.
{}From the equations of motion\cite{RST92a}, one can show that
\[ \frac{dx^{-}}{dx^{+}}\Big|_{\mbox{\tiny A.H.}}
    = -\frac{\partial_{+}^{2}\phi}{\partial_{+}\partial_{-}\phi}\,
       \Big|_{\mbox{\tiny A.H.}}.\]
So, we must have $\partial_{+}\rho$, $\partial_{+}^{2}\phi$ and
$\partial_{+}\partial_{-}\phi$ at the intersection point
between the apparent horizon and the shell line.

For the semi-classical action (\ref{classical_cghs_action})
(\ref{liouville_action}),
the equations of motion are
\begin{eqnarray}
  \label{eom_of_semi-classical_cghs_phi}
  e^{-2\phi}(4\partial_{+}\phi\partial_{-}\phi
  -4\partial_{+}\partial_{-}\phi
  +2\partial_{+}\partial_{-}\rho +\lambda^{2}e^{2\rho}) =0,
\end{eqnarray}
\begin{eqnarray}
  \label{eom_of_semi-classical_cghs_rho}
  e^{-2\phi}(4\partial_{+}\phi\partial_{-}\phi
  -2\partial_{+}\partial_{-}\phi +\lambda^{2}e^{2\rho})
  +\kappa\partial_{+}\partial_{-}\rho =0.
\end{eqnarray}
On the shell line ($x^{+} = x^{+}_{0}$), we define
$\Sigma \equiv \partial_{+}\phi$, $\Xi \equiv \partial_{+}\rho$
and $w = e^{-2\phi} = e^{-2\rho} = -\lambda^{2}x^{+}_{0}x^{-}$,
which is the LDV solution.
Boundary conditions for $\phi$ and $\rho$ is that they asymptotically
coincide with the classical solution because quantum correction
would be negligible in the asymptotic region.

With these variables, (\ref{eom_of_semi-classical_cghs_phi})
and (\ref{eom_of_semi-classical_cghs_rho})
can be written as
\begin{eqnarray}
  \label{eom_in_Sigma_1}
  \left( 2w -\kappa \right)\partial_{w}\Xi
  -2w\partial_{w}\Sigma = 0,\\
  \label{eom_in_Sigma_2}
  \kappa \partial_{w}\Xi -2w \partial_{w}\Sigma
  -2\Sigma -\frac{1}{x^{+}_{0}} = 0,
\end{eqnarray}
where we have used
$ \partial_{-}\phi = \lambda^{2}x^{+}_{0}/2w,\;
  \partial_{-} = -\lambda^{2}x^{+}_{0}\partial_{w}.$
By eliminating $\partial_{w}\Xi$, we have
\begin{eqnarray}
  \label{eom_of_F_CGHS}
  A \Sigma' +\frac{1}{2}A' \Sigma = -\frac{1}{4x^{+}_{0}}A',
\end{eqnarray}
where $'$ denotes $\partial_{w}$, and $A$, $A'$ are defined as
$  A \equiv 4w (w- \kappa),\;
  A' = \partial_{w}A = 8w - 4\kappa.$
This differential equation can be solved easily.
With the boundary condition, we have
\begin{eqnarray}
  \label{Sigma_solution}
  \Sigma = -\frac{1}{2x^{+}_{0}}
  + \frac{M}{\lambda x^{+}_{0}}\frac{1}{\sqrt{A}}.
\end{eqnarray}
Because the apparent horizon is defined by
$\partial_{+}\phi = \Sigma = 0$, we can find the point
{}from which the apparent horizon starts.
Then we have
\begin{eqnarray}
  \label{solution_Sigma=0}
  A_{0} = \frac{2M}{\lambda},\; A'_{0}=4\sqrt{\kappa^{2}
    + \left(\frac{2M}{\lambda}\right)^{2}}
\end{eqnarray}
or in terms of $w$,
\begin{eqnarray}
  \label{solution_Sigma=0_w}
  w_{0} = \frac{\kappa}{2}
  +\frac{1}{2}\sqrt{\kappa^{2}+\left(\frac{2M}{\lambda}\right)^{2}}.
\end{eqnarray}
We can find other solutions which satisfy (\ref{solution_Sigma=0}).
They are, however, not interesting because they are behind singularity.
More details will be mentioned later.

With the above solution of $\Sigma$, $\Xi$ is obtained
by integrating equation (\ref{eom_in_Sigma_2}),
\begin{eqnarray}
  \label{Xi_solution}
  \Xi = \frac{2w}{\kappa}\Sigma +\frac{A'}{8\kappa x^{+}_{0}}
  -\frac{M}{\kappa\lambda x^{+}_{0}}.
\end{eqnarray}
Using these results, we have
\begin{eqnarray}
  \label{partial_plus_S}
  \partial_{+}S|_{\mbox{\tiny A.H.}} =\frac{1}{x^{+}_{0}}
  \left\{ \kappa - \frac{2M}{\lambda}
  +\sqrt{\kappa^{2}+ \left(\frac{2M}{\lambda}\right)^{2}} \right\}.
\end{eqnarray}

Next, we calculate the other part of $dS/dx^{+}$.
The gradient of the apparent horizon contains
$\partial_{+}\partial_{-}\phi$ and $\partial_{+}^{2}\phi$,
the former can be estimated from the equations of motion directly.
Eliminating $\partial_{+}\partial_{-}\rho$ from
(\ref{eom_of_semi-classical_cghs_phi}) and
(\ref{eom_of_semi-classical_cghs_rho})
one has the value of $\partial_{+}\partial_{-}\phi$
on the apparent horizon as
\begin{eqnarray}
  \label{d_plus_minus_phi_AH}
  \partial_{+}\partial_{-}\phi|_{\mbox{\tiny A.H.}}
   = \frac{\lambda^{4}}{16 M^{2}}A'_{0}.
\end{eqnarray}
Now, we define a new variable ${\cal F} \equiv \partial_{+}^{2}\phi$.
Differentiating (\ref{eom_of_semi-classical_cghs_phi})
and (\ref{eom_of_semi-classical_cghs_rho})
and eliminating $\partial_{+}^{2}\rho$, we have the differential
equation for ${\cal F}$,
\begin{eqnarray}
  \label{equation_for_F}
  A{\cal F}' + \frac{1}{2}A'{\cal F}
   = A' \partial_{w}(w \Sigma^{2}) -\frac{A'}{2 x^{+}_{0}}(\Xi-\Sigma)
    + 4\kappa w \Sigma \partial_{w}(\Xi-\Sigma).
\end{eqnarray}
Solution of this differential equation with boundary condition is
\begin{eqnarray}
  \label{solutio_F}
  {\cal F}
  &=& \frac{1}{2 x^{+^{2}}_{0}}
    + \frac{M}{\kappa\lambda x^{+^{2}}_{0}}
    - \frac{M}{4\kappa\lambda x^{+^{2}}_{0}}\frac{A'}{\sqrt{A}}
    +\frac{M^{2}}{32\kappa\lambda^{2}x^{+^{2}}_{0}}
    \frac{A'^{3}}{A^{2}}\nonumber\\ 
  & & - \frac{M^{2}}{2\kappa\lambda^{2} x^{+^{2}}_{0}}\frac{A'}{A}
    -\frac{2M}{\lambda x^{+^{2}}_{0}}
    +\frac{2M^{2}}{32\kappa\lambda^{2}x^{+^{2}}_{0}}\frac{1}{A}
    -\frac{3\kappa M}{8\lambda x^{+^{2}}_{0}}\frac{A'}{A^{3/2}}\nonumber\\
  & & -\frac{3\kappa^{2}M}{2\lambda x^{+^{2}}_{0}}\frac{1}{A^{3/2}}
    +\frac{2\kappa^{2}M}{\lambda^{2}x^{+^{2}}_{0}}\frac{1}{A^{2}}
    +\frac{M}{4\lambda x^{+^{2}}_{0}}
    \log{\frac{A'+4\kappa}{A'-4\kappa}}.
\end{eqnarray}
Substituting $A_{0}$ and $A'_{0}$ into above result,
one finds the value of ${\cal F}$ at the apparent horizon,
and finally we have
\begin{eqnarray}
  \label{entropy_evolution}
    \frac{dS}{d x^{+}}\Big|_{\mbox{\tiny A.H.}}
 &=& \frac{2\kappa}{x^{+}_{0}}
 +\frac{1}{x^{+}_{0}}\left\{\sqrt{\kappa^{2}
     +\left(\frac{2M}{\lambda}\right)^{2}}-\frac{2M}{\lambda}\right\}
 + \frac{\kappa}{2x^{+}_{0}}\left\{1-
   \frac{\kappa}{\sqrt{\kappa^{2}+\left(\frac{2M}{\lambda}\right)^{2}}}
   \right\}\nonumber\\
 & & + \frac{\kappa^{2}\lambda^{2}}{4M^{2}}
 \left[ 1-
   \frac{\left(\frac{2M}{\lambda}\right)^{2}}
        {\sqrt{\kappa^{2}
            +\left(\frac{2M}{\lambda}\right)^{2}}
          \left\{
            \left(\frac{2M}{\lambda}\right)^{2}
            +\sqrt{\kappa^{2}+\left(\frac{2M}{\lambda}\right)^{2}
              }\right\}}
      \right]\nonumber\\
 & & +\frac{4M^{2}}{\lambda^{2}x^{+}_{0}}
 \log{\frac{\sqrt{\kappa^{2}+\left(\frac{2M}{\lambda}\right)^{2}}+\kappa}
   {\sqrt{\kappa^{2}+\left(\frac{2M}{\lambda}\right)^{2}}-\kappa}}.
\end{eqnarray}
It is easily seen that the last log term is always positive.
It can be proved that the
sum of the rest terms are always positive.
Therefore we could show that
the entropy does not decrease along the apparent horizon
at least in the vicinity of the infalling shell.

\subsubsection{Numerical approach}

To know the behavior of entropy along the apparent horizon
far from the shell,
we must perform numerical calculation for some values of the parameters
$\kappa$ and $M$.
Although we cannot cover all values of the parameters
in numerical calculation,
it is necessary to know
information which cannot be obtained from
the above analytic approach.

Numerical calculation was performed on $\sigma^{\pm}$ coordinate,
and we fix the parameters,
$\lambda = 1,\; \sigma^{+}_{0} = 0$ ($x^{+}_{0} = 1$)\cite{PS93}.
Equations for $\phi$ and $\rho$ are second order partial differential
equations. 
We have used the numerical schemes invented in \cite{HS93}.
The range of $\sigma^{+}$ is from the infalling shell
to the merging point of the apparent horizon and singularity.
One of the typical numerical results is shown in Figure 2.
The result is for $\kappa = M = 1$.
Solid curve is numerically calculated entropy, and dashed strait
line is analytically calculated one.
In the latter one, the value of entropy on the infalling shell is evaluated
by using the LDV solution, and gradient of entropy is calculated
with analytic scheme presented in above subsection.
This result shows that the entropy grows still even in the
region far from the infalling shell.
The range in which analytic approximation is good is not so wide
in this case, it shows the higher order effects make the
growing rate more rapidly.
We also calculated the evolution of the entropy for other parameters.
All of the results show the same feature.
\begin{figure}[t]
  \begin{center}
    \leavevmode
    \epsfile{file=fig2,width=12cm}
    \begin{small}
    \caption{This figure shows the entropy of semi-classical
      CGHS model for parameters
      $\lambda=1,\; \sigma^{+}_{0}=0\,(x^{+}_{0}=1),\; \kappa = M = 1$.
      Solid curve is numerically calculated entropy, and dashed strait
      line is analytically calculated one.}
    \end{small}
    \label{fig:entropy}
  \end{center}
\end{figure}

\section{Entropy of generic 2 dimensional model}
\label{sec:general_2_dimentional_model}

\subsection{Generalized action and entropy}
\label{sec:generalized_action_and_entropy}

The semi-classical CGHS action is given by adding the non-local
counter term $I_{1}$.
Other covariant local counter terms can be added to the action
because of ambiguity in defining measures used in the
functional integral, and combinations of these counter terms
make different models \cite{Myers95}.
Possible local counter terms which can be added to the action
are
\begin{eqnarray}
  \label{local_counter_term_2}
 I_{2} = -\frac{\kappa}{8\pi}\int d^{2}x \sqrt{-g}\{\alpha \phi R
  + \beta (\nabla \phi)^{2}\},
\end{eqnarray}
\begin{eqnarray}
  \label{local_counter_term_3}
 I_{3} = -\frac{\kappa}{8\pi}\int d^{2}x \sqrt{-g}
              \sum^{K}_{n=2}\{a_{n}\phi^{n}R 
              + b_{n}\phi^{n-1}(\nabla \phi)^{2}\}.
\end{eqnarray}
Because we will consider black hole solutions in these models,
which are asymptotically flat,
it is reasonable to require that models include
the LDV solution.
This requirement leads $a_{n} = b_{n} = 0$, so the
action $I_{3}$ will not be considered below.
If $1-\frac{\alpha}{2}-\frac{\beta}{4} = 0$, the model is exactly
soluble because of the symmetry which leads
$\partial_{+}\partial_{-}(\rho - \phi)=0$ just like the case of
classical CGHS model \cite{Myers95}.

Equations of motion derived from the generalized action
$I = I_{0}+I_{1}+I_{2}$ are
\begin{eqnarray}
  \label{generalized_eom_phi}
  e^{-2\phi}(4\partial_{+}\phi\,\partial_{-}\phi
  - 4\partial_{+}\partial_{-}\phi +2\partial_{+}\partial_{-}\rho
  + \lambda^{2}e^{\,2\rho})
  + \frac{\kappa}{4}(\alpha \partial_{+}\partial_{-}\rho
  + \beta\partial_{+}\partial_{-}\phi) = 0,
\end{eqnarray}
\begin{eqnarray}
  \label{generalized_eom_rho}
   e^{-2\phi}(4\partial_{+}\phi\,\partial_{-}\phi
          -2 \partial_{+}\partial_{-}\phi + \lambda^{2}e^{\,2\rho})
          +\kappa (\partial_{+}\partial_{-}\rho
          -\frac{\alpha}{4}\partial_{+}\partial_{-}\phi)=0,
\end{eqnarray}
and the constraints are
\begin{eqnarray}
  \label{generalized_constraints}
   & & e^{-2\phi}(2\partial_{\pm}^{2}\phi
     -4\partial_{\pm}\phi\,\partial_{\pm}\rho)
     -\frac{1}{2}\sum_{i=1}^{N}\partial_{\pm}f_{i}
     \partial_{\pm}f_{i}\nonumber\\
             & & +\kappa \{(\partial_{\pm}\rho)^{2}
     -\partial_{\pm}^{2}\rho + t_{\pm} - \frac{\alpha}{4}
     (2\partial_{\pm}\phi\,\partial_{\pm}\rho
     -\partial_{\pm}^{2}\phi) - \frac{\beta}{4}
     (\partial_{\pm}\phi)^{2}\}  =0.
\end{eqnarray}
The determinant of the coefficient matrix of the kinetic term is
\begin{equation}
  \left\{2 e^{-2\phi} + \kappa\left(\frac{\alpha}{4}-1\right)\right\}^2 
  -\kappa^{2}\left(1-\frac{\alpha}{2}-\frac{\beta}{4}\right).
\end{equation}
In the case $ 1- \alpha / 2 -\beta /4 < 0$,
there is no degeneracy because the above formula always takes positive
value.
It means that the solution has no singularity.
As we would like to consider the entropy of the evaporating black holes,
this case is excluded from our consideration.
By substituting the LDV solution (\ref{vacuum_solution})
into the constraint (\ref{generalized_constraints}),
the boundary terms are determined as
\begin{eqnarray}
  \label{boundary_terms}
 t_{\pm} = \frac{1}{4(x^{\pm})^{2}}(1+ \frac{\beta}{4}).
\end{eqnarray}
If $\beta=0$, this corresponds to the boundary conditions
adopted in the CGHS model and RST model.
We then have
\begin{equation}
  <T_{\pm\pm}> = -{\beta \over 4}{\kappa \lambda^2 \over 4},
\end{equation}
which are evaluated from the Liouville part of the
constraints (\ref{generalized_constraints}).
This means that the LDV spacetime is filled with radiation\cite{BPP95}.
We consider only the case $\beta<0$ as physical system.

\subsection{Analytic approach for generic models}
\label{sec:analytic_approach_for_generic_models}

With these general action and equations, we can also investigate
the behavior of entropy in generic model.
Extra term of entropy which did not appeared in CGHS model is
\begin{eqnarray}
  \label{S_2}
  S_{2} = -\frac{\kappa\alpha}{2}\phi,
\end{eqnarray}
that comes from the local counter term $I_{2}$.
Thus the resulting formula for the entropy is given by
\begin{eqnarray}
  \label{S_general}
  S = 2e^{-2\phi} + 2\kappa\rho -\frac{\kappa\alpha}{2}\phi
  + \kappa\log(-\lambda^{2}x^{+}x^{-}).
\end{eqnarray}

Calculation of the entropy evolution near the in-falling
infalling shell is almost similar to the
case of semi-classical CGHS model.
Equations of $\Sigma$ and $\Xi$ are
\begin{eqnarray}
  \label{eom_Sigma_generic_1}
  \left\{2w + \kappa\left(\frac{\alpha}{4}-1\right)\right\}\partial_{w}\Xi
  -\left\{2w-\frac{\kappa}{4}(\alpha+\beta)\right\}\partial_{w}\Sigma
  = 0,\\
  \kappa\partial_{w}\Xi -\left(2w+\frac{\alpha\kappa}{4}\right)
  \partial_{w}\Sigma -2\Sigma -\frac{1}{x^{+}_{0}} = 0.
\end{eqnarray}
The solutions are
\begin{eqnarray}
  \label{Sigma_solution_generic}
  \Sigma = -\frac{1}{2x^{+}_{0}}+\frac{M}{\lambda x^{+}_{0}}
  \frac{1}{\sqrt{B}},
\end{eqnarray}
and
\begin{eqnarray}
  \label{Xi_solution_generic}
  \Xi = \frac{1}{4\kappa}(B'+4\kappa)\Sigma + \frac{1}{8\kappa x^{+}_{0}}B'
  -\frac{M}{\lambda\kappa x^{+}_{0}},
\end{eqnarray}
where $B$ and $B'$ are defined to be
\begin{eqnarray}
  \label{def_B_generic}
  B &=& \left\{2w +\kappa\left(\frac{\alpha}{4}-1\right)\right\}^{2}
  -\kappa^{2}\left(1-\frac{\alpha}{2}-\frac{\beta}{4}\right),\nonumber\\
  B'&=& 4\left\{2w +\kappa\left(\frac{\alpha}{4}-1\right)\right\},
\end{eqnarray}
and these can be reduced into $A$ and $A'$ if $\alpha = \beta = 0$.

Solving $\Sigma=0$, one obtains the point from which apparent horizon starts.
For $1-\alpha/2 -\beta/4>0$, there are two solutions.
We take the largest one as a solution, $w_{0}$,
because the other one is hidden behind of the singularity.
By solving $\Sigma=0$,
\begin{eqnarray}
  \label{solution_B}
  B_{0} &=& \left( \frac{2M}{\lambda} \right),\nonumber\\
  B'_{0}&=& 4\sqrt{\kappa^{2}\left( 1-\frac{\alpha}{2}-\frac{\beta}{4}\right)
    +\left( \frac{2M}{\lambda}\right)^{2}}.
\end{eqnarray}
With the above results, contribution of $\partial_{+}S$
at apparent horizon can be evaluated.

Contribution of $\partial_{-}S$ can be calculated in the
same way to CGHS model.
Differential equation for $\cal F$ has also the same form as
(\ref{equation_for_F}),
\begin{eqnarray}
  \label{equation_for_F_with_B}
  B{\cal F}' + \frac{1}{2}B'{\cal F}
   = B' \partial_{w}(w \Sigma^{2}) -\frac{B'}{2 x^{+}_{0}}(\Xi-\Sigma)
    + 4\kappa w \Sigma \partial_{w}(\Xi-\Sigma).
\end{eqnarray}
We must consider two cases separately, one is $1-\alpha/2 -\beta/4 = 0$,
another case is $1-\alpha/2-\beta/4 >0$.
Semi-classical CGHS model belongs to the latter case.

\subsubsection{Exactly soluble case}

In the case of $1-\alpha/2-\beta/4 = 0$, which is exactly soluble
like RST model\cite{RST92b} or model of Bose,{\it et.al.\/}
\cite{BPP95},
we can take $\phi=\rho$ gauge, so that the latter two terms of
(\ref{equation_for_F_with_B}) are zero.
Thus the solution and the value at the apparent horizon
can be found easily.
Then we have the time evolution of entropy in the exactly
soluble case as
\begin{eqnarray}
  \label{d_plus_S_soluble_case}
  \frac{dS}{dx^{+}}\Bigm|_{\mbox{\tiny A.H.}}
 = \frac{\kappa\alpha}{4x^{+}_{0}}
 \frac{\frac{2M}{\lambda}}{\frac{2M}{\lambda}
   -\kappa\left(\frac{\alpha}{4}-1\right)}.
\end{eqnarray}
This takes positive value when model-fixing parameter satisfies
\begin{eqnarray}
  \label{alpha_range_soluble_case}
  0 < \alpha < 4\left(\frac{2M}{\kappa\lambda}+1\right),
\end{eqnarray}
or equivalently,
\begin{eqnarray}
  \label{beta_range_soluble_case}
  -4\left(\frac{4M}{\kappa\lambda} +1 \right) < \beta < 4.
\end{eqnarray}
In the case of RST model ($\alpha=2,\; \beta=0$)\cite{Myers94},
the increasing rate of entropy is
\begin{eqnarray}
  \label{case_of_RST}
    \frac{dS}{dx^{+}}\Bigm|_{\mbox{\tiny A.H.}}
 = \frac{\kappa}{2x^{+}_{0}}\frac{\frac{2M}{\lambda}}
 {\frac{2M}{\lambda}+\frac{\kappa}{2}} >0,
\end{eqnarray}
and in the case of Bose,{\it et.al\/} model ($\alpha=4,\;\beta=-4$),
\begin{eqnarray}
  \label{case_of_BPP}
    \frac{dS}{dx^{+}}\Bigm|_{\mbox{\tiny A.H.}}
    = \frac{\kappa}{x^{+}_{0}} >0.
\end{eqnarray}

\subsubsection{Unsoluble case}
In the other case, $1-\frac{\alpha}{2}-\frac{\beta}{4} >0$,
the solution of (\ref{equation_for_F_with_B}) cannot be reduced into
one of exactly soluble case, so we must distinguish this case
{}from the previous one.
With the solution of (\ref{equation_for_F_with_B}) and contribution of
$\partial_{+}S$, we have
\begin{eqnarray}
  \label{d_plus_S_general}
 \frac{dS}{d x^{+}}\Big|_{\mbox{\tiny A.H.}}
 &=& \frac{1}{x^{+}_{0}}
 \left\{\kappa -\frac{2M}{\lambda}
   +\sqrt{\kappa^{2}\left(1-\frac{\alpha}{2}-\frac{\beta}{4}\right)
     +\left(\frac{2M}{\lambda}\right)^{2}}\right\}\\
 & & +\frac{B'_{0} +4\kappa}{B'_{0}-4\kappa \left(\frac{\alpha}{4}-1\right)}
 \Biggl[\;
       \frac{\kappa}{2x^{+}_{0}}
       \left\{3-2\left(\frac{\alpha}{4}-1\right)\right\}
       - \frac{4M^{2}}{\kappa\lambda^{2}x^{+}_{0}}\nonumber\\
 & & + \frac{32M^{3}}{\kappa\lambda^{3}x^{+}_{0}}\frac{1}{B'_{0}}
       -\frac{2\kappa^{2}}{x^{+}_{0}}
       \left(1-\frac{\alpha}{2}-\frac{\beta}{4}\right)
       \frac{1}{B'_{0}}\nonumber\\
 & & +\frac{4M^{2}}{\lambda^{2}x^{+}_{0}}
      \frac{2\left(1-\frac{\alpha}{2}-\frac{\beta}{4}\right)
        + \frac{\alpha}{4}-1}
      {\sqrt{1-\frac{\alpha}{2}-\frac{\beta}{4}}}
      \log{\frac{B_{0}'+4\kappa\sqrt{1-\frac{\alpha}{2}-\frac{\beta}{4}}}
        {B_{0}'-4\kappa\sqrt{1-\frac{\alpha}{2}-\frac{\beta}{4}}}}\;
      \Biggr].
\end{eqnarray}
The sign of this value depends on $\alpha$ and $\beta$.
For example, in the case $\alpha=12,\; \beta=-20.1$
at $\lambda=1,\;x^{+}_{0}=1$,
which has slightly different parameters from that of the exactly soluble case,
$dS/dx^{+}$ takes the negative sign,
leads also negative $dS/dx^{+}$ near $\kappa=M$.
Therefore there exist some models which
have decreasing entropy with Wald's scheme in 2 dimension.

\section{Conclusion}
\label{sec:conclusion}

In this paper, we have investigated the dynamical evolution
of the entropy proposed by Wald
along the apparent horizon
in the context of the 2-dimensional dilaton gravitational 
theory.
In particular, we have considered the quantum backreaction
and then the evaporating spacetimes.
In the case of the semi-classical CGHS model,
we have shown analytically that the thermodynamic second
law holds along the apparent horizon in the vicinity of the
infalling shell.
Furthermore, the numerical calculation is also presented,
which suggest that the thermodynamic second law holds
even in the region far from the infalling shell.
Myers's result on the RST model and analysis on
the semi-classical CGHS model are very impressive,
however, they are yet special models.
Hence, we have also studied the general models which seem
physically natural in our judge.
Analytic calculation shows that, even in the vicinity of the
infalling shell, the thermodynamic law does not hold in
certain models.
There are two possible choices here.
One is to use this fact for restrictions of models.
Another is to modify the entropy formula.
In the latter case, the statistical interpretation of the
entropy would be necessary.
However, this interesting subject is beyond the scope of
this paper.
Further implications of our analysis should be considered
in the future.

\section*{Acknowledgments}
 The work is supported by Monbusho Grant-in-Aid
for Scientific Research No.~07740216.



\begin{thebibliography}{99}
\bibitem{Hawking73}
J.M.Bardeen, B.Carter, and S.W. Hawking,
{\sl Comm.\ Math.\ Phys.\/ \bf 31} (1973) 161.

\bibitem{Hawking75}
S.W. Hawking,
{\sl Comm.\ Math.\ Phys.\/ \bf 43} (1975) 199.

\bibitem{CGHS}
C.G. Callan, S.B. Giddings, J.A. Harvey, and A. Strominger,
{\sl Phys.\ Rev.\ D\/\bf 45} (1992) R1005.

\bibitem{frolov}
V.P.Frolov,
{\sl Phys.\ Rev.\ D\/\bf 46} (1992) 5383.

\bibitem{Wald93}
R.M. Wald,
{\sl Phys.\ Rev.\ D\/\bf 48} (1993) R3427.

\bibitem{JKM}
T.Jacobson, G.Kang, and R.C.Myers,
{\sl Phys.\ Rev.\ D\/\bf 49} (1994) 6587.

\bibitem{IyerWald} V. Iyer and R. M. Wald,
{\sl Phys.\ Rev.\ D\/\bf 50} (1994) 846.

\bibitem{Myers94}
R.C. Myers,
{\sl Phys.\ Rev.\ D\/\bf 50} (1994) 6412.

\bibitem{RST92b}
J.G. Russo, L. Susskind and L. Thorlacius,
{\sl Phys.\ Rev.\ D\/\bf 46} (1992) 3444.

\bibitem{tada}
T.Tada and S.Uehara,
{\sl Phys.\ Lett.\ B\/ \bf 305} (1993) 23.

\bibitem{RST92a}
J.G. Russo, L. Susskind and L. Thorlacius,
{\sl Phys.\ Lett.\ B\/ \bf 292} (1992) 13.

\bibitem{PS93}
T. Piran and A. Strominger,
{\sl Phys.\ Rev.\ D\/\bf 48} (1993) 4729.

\bibitem{HS93}
S.W. Hawking and J.M. Stewart,
{\sl Nucl.\ Phys.\ B\/ \bf400} (1993) 393.

\bibitem{Myers95}
G. Michaud and R.C. Myers,
gr-qc/9508063.

\bibitem{BPP95}
S. Bose, L. Parker, and Y. Peleg,
{\sl Phys.\ Rev,\ D\/\bf 52} (1995) 3512.

\end{thebibliography}
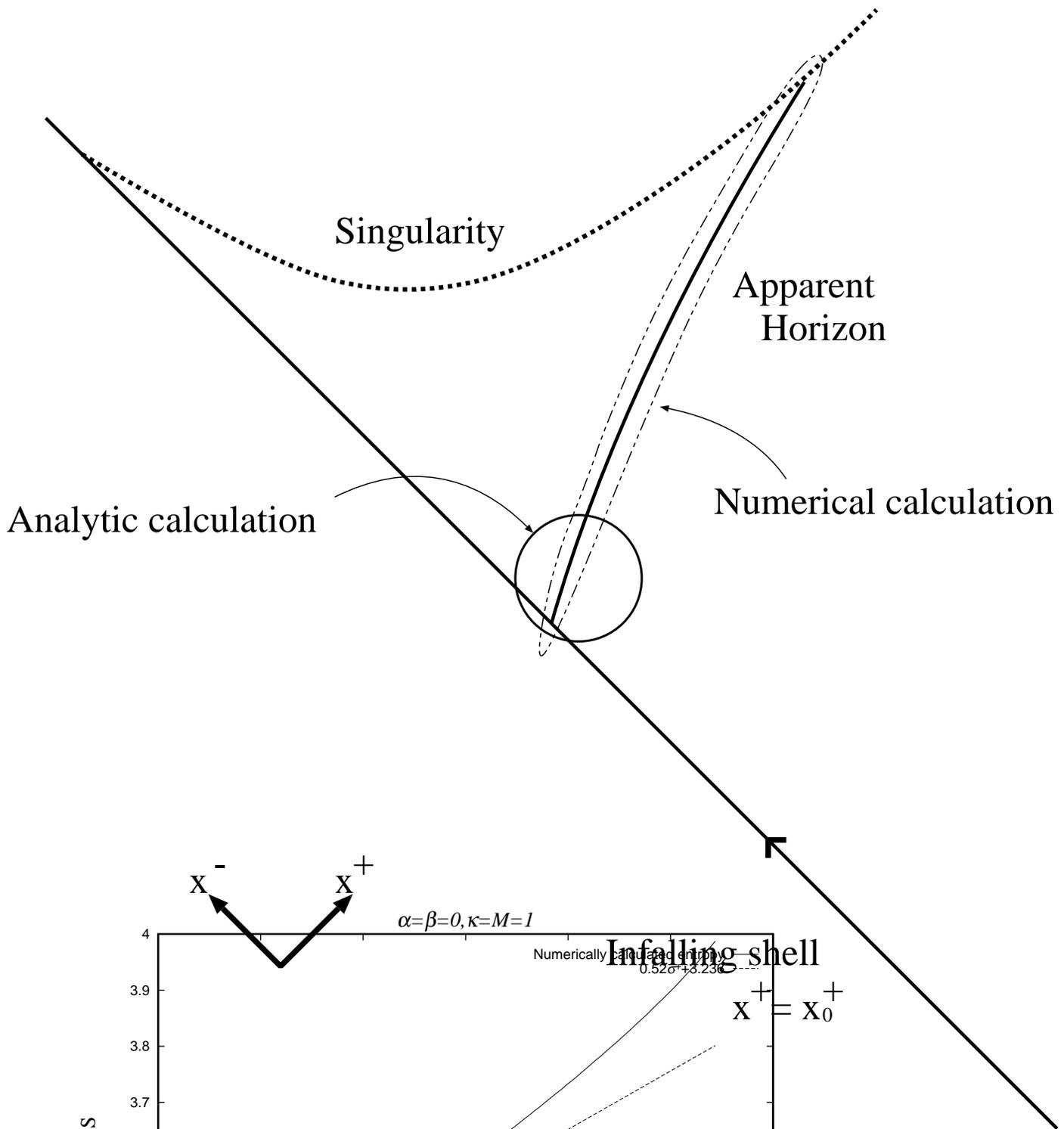
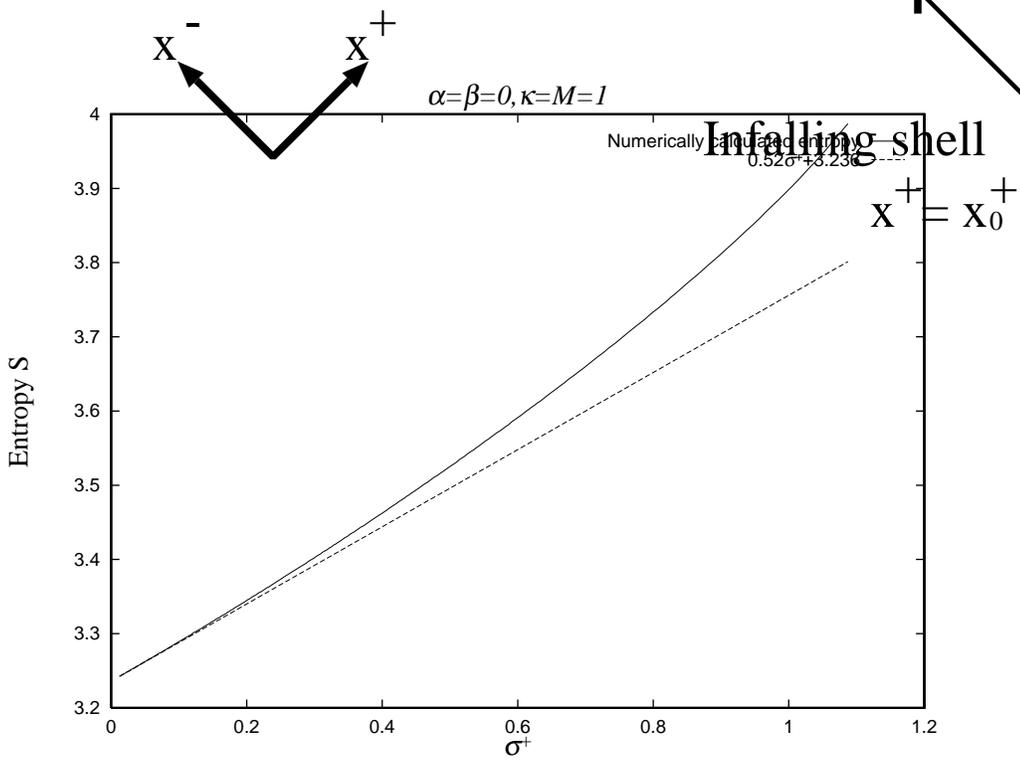
\end{document}